%% file: mcsim_CRcomp.tex
\makeatletter
\newcommand{\figcaption}[1]{\def\@captype{figure}\caption{#1}}
\newcommand{\tblcaption}[1]{\def\@captype{table}\caption{#1}}
\makeatother

\documentclass{PoS}
\usepackage{graphicx}
\usepackage{indentfirst}
\usepackage{wrapfig}
\title{A Monte Carlo simulation study for cosmic-ray chemical composition measurement with Cherenkov Telescope Array}
\ShortTitle{A MC simulation study for CR chemical composition measurement with CTA}

\author{\speaker{Michiko Ohishi}$^1$, Takanori Yoshikoshi$^1$, Tatsuo Yoshida$^2$ for the CTA Consortium\\
\llap{$^1$}Institute for Cosmic Ray Research, the University of Tokyo, Chiba 277-8582, Japan\\
\llap{$^2$}College of Science, Ibaraki University, Ibaraki 310-8512, Japan \\
E-mail: \email{ohishi@icrr.u-tokyo.ac.jp}
}

\abstract{Our Galaxy is filled with cosmic-ray particles and more than 98\% of them  are atomic nuclei. In order to clarify their origin and acceleration  mechanism, chemical composition measurements of these cosmic rays with  wide energy coverage play an important role.

   Imaging Atmospheric Cherenkov Telescope (IACT) arrays are designed to  detect cosmic gamma-rays in the very-high-energy regime ($\sim$TeV). Recently these systems proved to be capable of measuring cosmic-ray chemical composition in the sub-PeV region by capturing direct Cherenkov photons emitted by charged primary particles. Extensive air shower profiles measured by IACTs also contain information about the primary particle type since the cross section of inelastic scattering in the air depends on the primary mass number.

   The Cherenkov Telescope Array (CTA) is the next generation IACT system, which will consist of multiple types of telescopes and have a km$^2$-scale footprint and extended energy coverage (20 GeV to 300 TeV). In order to estimate CTA potential for cosmic ray composition measurement, a full Monte Carlo simulation including a description of extensive air shower and detector response is needed.
  We generated a number of cosmic-ray nuclei events (8 types selected from H to Fe) for a specific CTA layout candidate in the 
southern-hemisphere site. We applied Direct Cherenkov event selection and shower profile analysis to these data and preliminary results on charge number resolution and expected event count rate for these cosmic-ray nuclei are presented.
}

\FullConference{35th International Cosmic Ray Conference - ICRC2017 -\\
		10-20 July, 2017\\
		Bexco, Busan, Korea}

\begin{document}

\section{Introduction}
\input{sec1.tex}

\section{Monte Carlo Simulation Setup}
\input{sec2.tex}

\section{Analysis}

\input{sec30.tex}

\subsection{Direct Cherenkov analysis}
\input{sec31.tex}

\subsection{Shower parameter Multivariate Analysis (MVA) }
\input{sec32.tex}

\subsection{Comparison of results from two methods, two hadron interaction models}
\input{sec33.tex}

\section{Summary and future plan}
\input{sec4.tex}
\acknowledgments
 This work was conducted in the context of the CTA Analysis and Simulation Working Group/Physics Cosmic-Ray Science Working Group. We gratefully acknowledge financial support from the agencies and organizations listed here: \verb|http://www.cta-observatory.org/consortium_acknowledgments|. This work was supported by Grant-in-Aid for Scientific Research No. 26800126 from Japan Society for the Promotion of Science.

\end{document}

%% file: sec1.tex
   Our Galaxy is filled with high energy cosmic rays and a large majority (>98\%) of them consists of atomic nuclei. Since the characteristics of acceleration and propagation of a particle depend on its charge and mass, cosmic-ray chemical composition measurements over wide energy range are important to understand the nature of cosmic-ray origin and propagation.

   Cosmic-ray composition measurements in TeV-PeV region are carried out by balloons/satellites and air shower detectors on the ground. The direct measurements by balloons and satellites achieve good charge resolution but suffer from poor event statistics because of small ($\sim$ m$^2$) effective area. The indirect measurements from the ground benefit from large collection area ($\sim$ km$^2$) but with relatively poor charge resolution since they detect only secondary particles. Thus cosmic-ray composition measurement in this energy region is still a difficult task and measurement results from various experiments show a wide variety as shown in Fig.9 of \cite{horandel2}.

              Imaging Atmospheric Cherenkov Telescopes (IACTs) are designed to detect very-high-energy ($\sim$ TeV) cosmic gamma-rays. Recently, these systems proved to be capable of measuring cosmic-ray chemical composition in sub-PeV region by using direct Cherenkov photons emitted by charged primary particle. On the basic idea proposed by Kieda et al. \cite{Kieda},  H.E.S.S. and VERITAS obtained cosmic-ray iron (Z=26) spectra using this method in 13-200 TeV \cite{HESS} and 20-500 TeV \cite{VERITAS} region, respectively. Shower profile measured by IACTs also contains information of primary particle type since the cross section of the first interaction depends on the primary mass number.

In the IACT field, a major next-generation project Cherenkov Telescope Array (CTA) is currently in the construction phase. The CTA array will cover a  km$^2$-scale footprint with multiple types of telescopes and is expected to achieve higher event statistics in cosmic-ray nuclei measurement than the current systems. In order to assess the capability of CTA in the cosmic-ray chemical composition measurement, we performed a Monte Carlo simulation which includes descriptions of extensive air shower and CTA detector response for various cosmic-ray nuclei inputs from proton (Z=1) to iron (Z=26). We describe the simulation setup in section 2, the analysis methods and results (charge resolution and expected count rates) in section 3.

%% file: sec2.tex
CTA Monte Carlo simulation tool (corsika{$\_$}simtelarray) consists of general-purpose air shower simulator CORSIKA\cite{CORSIKA} and CTA detector response description ({\it sim$\_$telarray}). Details about the simulation tool are described in \cite{read_cta}. We used the version called {\it Production 3} in this study.

\begin{figure}[t]
\begin{minipage}[b]{6cm}
    \includegraphics[width=6cm,keepaspectratio]{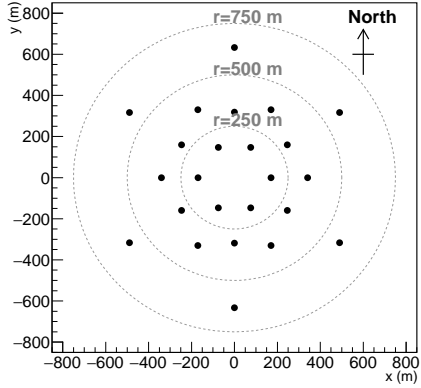}
    \figcaption{Array configuration used in the simulation. All the 24 telescopes are middle-sized telescopes (MSTs).}
        \label{fig:arrayconfig}

    \end{minipage}
     \hspace*{8mm}
   \begin{minipage}[b]{8cm}
        \small
   \begin{tabular}{l|c}
  \hline
  Parameter & Value \\
  \hline
  Site & Paranal, Chile \\
  Observation level & 2150 m \\
  Zenith angle & 20 deg  \\
  Azimuthal angle & 180 deg (from North) \\
  Shower core radius & 2000 m \\
  Viewcone & 0-10 deg \\
  Primary type & H, He, C, O, Ne, Mg, Si,Fe \\
  \hline
  \end{tabular}
 
    \tblcaption{Simulation parameters used in CORSIKA. Lower bound of simulated energy (TeV) is set to be 1.0 for H and He, 1.4 for C and O, 1.7, 2.1, 2.5, 5.0 for Ne, Mg, Si and Fe, respectively. Upper bound is set to be 1000 TeV for all the nuclei.}
      \label{tab:corpara}

  \end{minipage}
  \end{figure}

 Among several candidates of array configuration, we selected a moderate spacing one called 3HB1-3 and removed all the small-sized telescopes (SSTs) from the simulation since their spacing is too large for direct Cherenkov analysis. Then we have 24 middle-sized telescopes (MSTs) and 4 large-sized telescopes (LSTs) in the data, but this paper only treats the results for 24 MSTs with focal plane instruments called NectarCam \cite{NectarCam}. The array configuration of 24 MSTs is shown in Fig.\ref{fig:arrayconfig}.

As for cosmic-ray inputs, we selected 8 types of relatively abundant nuclei from proton (Z=1) to iron (Z=26), shown in Tab.\ref{tab:corpara} together with other simulation parameters. The spectral index was set as $-2.0$ in the simulation and events were re-weighted to fit the literature index value (H\"orandel 2003) \cite{Horandel} in the analysis process. We generated order of $10^7$ events for each nucleus. In order to check the effect of uncertainty of hadron interaction model which is one of a major source of systematic errors, we tested two models (QGSJET-II-03 and SIBYLL2.1, both in CORSIKA 6.990) in the event generation.

%% file: sec30.tex
 The simulated cosmic-ray event data were processed with signal integration and image cleaning as a similar approach in gamma-ray analysis. We set integration window to be 8 ns around signal peak of each pixel and adopted two-level tail-cut image cleaning \cite{read_cta} with relatively high pixel threshold (75 and 150 p.e.). Since we treat very energetic ($>$ 10 TeV) hadron events, high threshold helps to remove accompanying muon rings and extract bright shower cascade structure. This cleaning condition results in an energy threshold of $\sim 15$ TeV, angular resolution of $\sim0^\circ.1$ and core position resolution of $\sim 10$ m for iron events. Reconstruction of arrival direction/energy and calculation of basic image parameters were performed using $read\_cta$ \cite{read_cta} tool.

%% file: sec31.tex
  Direct Cherenkov (DC) analysis in this paper basically follows the H.E.S.S analysis \cite{HESS}. One of the unique feature of DC event is the existence of a single bright pixel (DC pixel) between shower image and arrival direction and we searched for such events using {\it DC-ratio} defined as:
\[ Q_{\rm DC}=\frac{I_{\rm max\_neighb.}}{I_{\rm pixel}} \]
  where $I_{\rm max\_neighb.}$ is the maximum intensity of the neighboring pixels. The analysis parameters used in the DC event search are shown in Tab.\ref{tab:DCsel}. Though Cherenkov photon arrival timing is also thought to be useful in the identification of DC events, we did not consider it since MST reflector is not isochronous which results in smearing of the signal.
  
 After the selection of DC candidate events, primary charge was reconstructed from the signal amplitude of DC pixel after subtracting average of signal amplitude of neighboring pixels ($I_{\rm DC}$). In the charge reconstruction process we used a conversion factor table made from {\it pure} iron direct Cherenkov events, where primary particles do not generate extensive air showers by controlling first interaction height in CORSIKA. The conversion factor depends on 4 parameters (primary energy $E$, impact parameter $r_{\rm impact}$, offset angle $\theta_{\rm off}$ and first interaction height), and the first 3 parameters can be reconstructed event by event. As for the last one we used the average for the events which survived the DC event selection. Thus reconstructed charge of a single DC candidate image is obtained as $ Z_{rec}=26\sqrt{I_{\rm DC}/C(E,r_{\rm impact},\theta_{\rm off})}$, where $C(E,r_{\rm impact},\theta_{\rm off})$ is taken from the conversion factor table. Then we took average of this value over telescopes (we required at least two telescopes have good DC candidate events for the charge reconstruction).
 \begin{table}[t]
 \small
 \begin{minipage}[t]{7.5cm}
    \begin{tabular}{c|l}
      \hline\hline
      Parameter & range \\
      \hline
      $\Delta^{\rm c.o.g}_{\rm DC} $(deg) & 0.17 to 0.91  \\
      $\Delta^{\perp}_{\rm DC} $(deg)& <0.20  \\
      $\Delta^{\rm dir}_{\rm DC} $ (deg) & <0.44  \\
     $I_{\rm DC-pixel}$ (p.e.) & < 4000  \\
      $ Q_{\rm DC}$ & < $0.223\log_{10}x(Size) - 0.246 $ \\
          $R_{\rm core}$ (m) & 40 to 140  \\
           $ N_{\rm tel}$ & $\geq 2$ \\
       $ Size $ (p. e.)  & < $10^5$  \\
       \hline
    \end{tabular}
      \tblcaption{Analysis parameters used in the selection of DC candidate events. Notation of the parameters is taken from \cite{HESS}.}
    \label{tab:DCsel}

  \end{minipage}\hspace*{6mm}
  \begin{minipage}[t]{7.5cm}
\begin{center}
\begin{tabular}{c|l}
     \hline \hline
     Parameter & range \\
      \hline
      $MSCW$ & -5 to 5 \\
      $X_{\rm max}^{\rm corr}$  (g/cm$^2$)& $< 600$ \\ 
      $MSCL$ & -5 to 5  \\
      $X_{\rm max}^{\rm err}/X_{\rm max} $ & $< 0.5$ \\
      $r_{\rm telmean}$ & $<500$  \\
      $E_{\rm resol}$ & $<0.2$  \\
      $E_{\chi^{2}}$ & $<0.5$  \\
 \hline       
    \end{tabular}
    \caption{Basic shower parameters used in the multivariate analysis (MVA) and their pre-cut values. The order of the parameters is same as the one of importance in MVA training process (31.6-100 TeV band).}
                   \label{tab:MVAprecut}
                   \end{center}
   \end{minipage}
\end{table}

\begin{figure}[b]
  \begin{center}
\includegraphics[width=12.5cm,keepaspectratio]{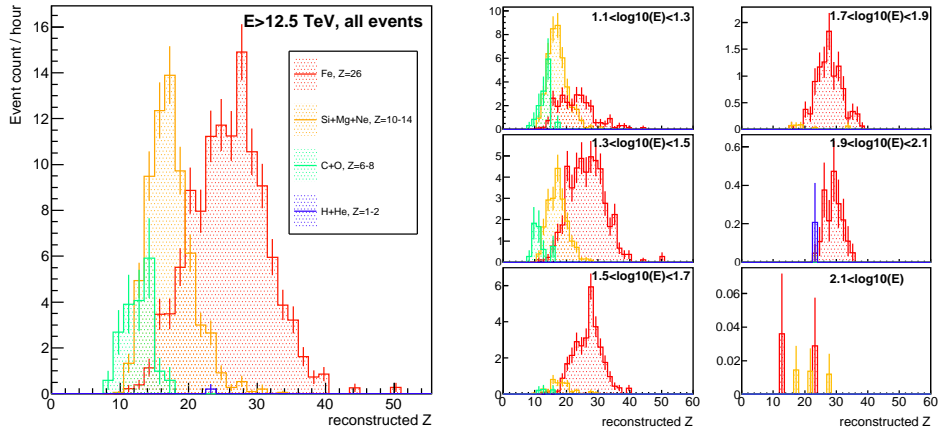}
      \caption{Reconstructed charge distributions obtained from DC analysis for 8 types of nuclei (preliminary value). Left: All events for E>12.5 TeV, Right: events divided into 6 energy bins. Since number of events of light nuclei is poor, they are shown as sum in groups (Si+Mg+Ne, C+O, and H+He). Only results for QGSJET-II-03 dataset are shown, which is common with Fig.2 to Fig.6.}
     \label{fig:DCrec}
    \end{center}
\end{figure}

The resulting reconstructed charge distributions for 8 types of nuclei are shown in Fig.\ref{fig:DCrec}. Events were weighted with cosmic-ray spectra from the literature \cite{Horandel}, thus the vertical axis corresponds to the expected event rate. Width of the reconstructed charge distribution (or charge resolution) of iron is found to be $4.33\pm0.20$  in $1.5<\log_{10}(E)<1.7$ energy bin (preliminary value and only statistical error is considered). Most of protons and heliums could not survive the DC event selection and remain as a small fraction (<1\% of iron) of background contamination.

%% file: sec32.tex
\begin{figure}[t]
  \begin{center}
     \includegraphics[width=13.5cm,keepaspectratio]{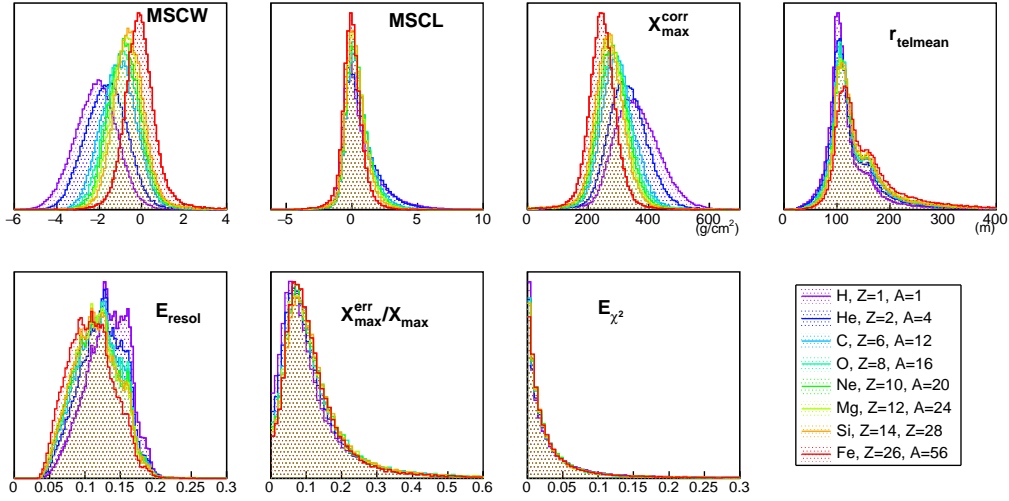}
     \caption{Distributions of basic shower parameters for 8 types of nuclei. In the calculation of scaled parameters, lookup-table made from iron events was used. $MSCW/L$: Mean Reduce Scaled WIDTH/LENGTH, X$_{\rm max}^{\rm corr}$: energy-dependence corrected X$_{\rm max}$, r$_{\rm telmean}$: average of telescope impact parameter weighted by $Size$ of the images, $E_{\rm resol}$: uncertainty of the reconstructed energy from the lookup table, $X_{\rm max}^{\rm err}$: Error in reconstructed X$_{\rm max}$, $E_{\chi^2}$: $\chi^2$ from the reconstructed energy to all telescope images. We applied pre-cut in these parameters for efficient MVA where the cut values are shown in Tab.\ref{tab:MVAprecut}.}
        \label{fig:showerparas}
\end{center}
\end{figure}

Shower parameters measured by IACTs also include information of primary type (there are previous studies such as \cite{Plya}), because difference in nucleon number leads to the difference in evolution of extensive air shower. For example, it is known that X$_{\rm max}$ (atmospheric depth where the number of particles in the shower gets maximum) parameter is useful in composition measurement. We checked distributions of seven basic shower parameters four our nuclei data, which are shown in Fig.\ref{fig:showerparas}. 

  We put these parameters into multivariate analysis (MVA) after rough pre-cut (cut values are shown in Tab.\ref{tab:MVAprecut}), in order to obtain a single index for nuclide identification. MVA is a popular method in IACT field since it is often used in gamma-hadron separation process. We chose a very simple MVA method, Fisher discriminant \cite{Fisher} (just a linear combination of the input parameters) and used ROOT TMVA\cite{TMVA} in the implementation.  Gaussian peak-like response of Fisher Discriminant was easy to handle in the estimation of charge number and resolution, but this old MVA method will be replaced more sophisticated one like Boosted Decision Tree (BDT) in the future.
  
Since the shower parameter distribution is energy dependent, MVA was trained in 3 different energy bands (12.5-31.6 TeV, 31.6-100 TeV and 100-1000 TeV), using iron nuclei as signal and protons as background to make separation of them maximum. We rescaled resulting MVA parameter so that iron peak comes at 26 and proton peak at 1 in order to treat this rescaled MVA parameter as charge number estimator.

\begin{figure}[t]
  \begin{center}
\includegraphics[width=12.5cm,keepaspectratio]{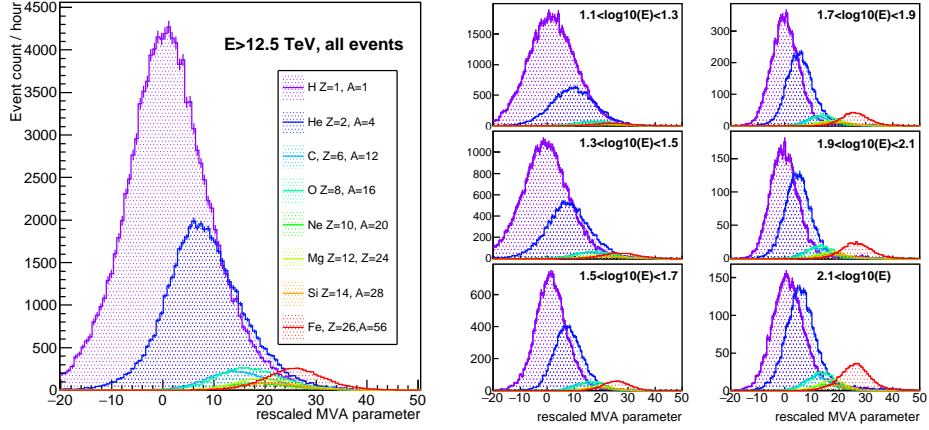}
      \caption{Resulting shower MVA parameter distributions, weighted by the flux value from the literature \cite{Horandel}. Left: All events for $E>12.5 $TeV, Right: events divided into 6 energy bin.}
     \label{fig:showerMVA_flux}
    
    \end{center}
\end{figure}

The resulting rescaled MVA parameter distributions for 8 types of nuclei are shown in Fig.\ref{fig:showerMVA_flux}, where events were weighted by the literature flux value \cite{Horandel}, as the same form of Fig.\ref{fig:DCrec}. The distribution width of charge is clearly energy dependent and resolution of iron is found to be $5.13\pm0.04$  in $1.5<\log_{10}(E)<1.7$ energy bin (preliminary value and only statistical error is considered).

%% file: sec33.tex
\begin{wrapfigure}[15]{r}[3mm]{65mm}
\centering
      \includegraphics[width=5.4cm,keepaspectratio]{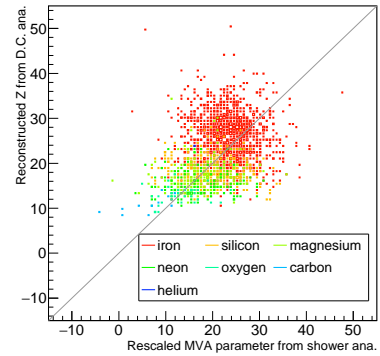}
     \caption{Relation between estimated charge numbers obtained from two different methods (Direct Cherenkov and shower MVA). All event for E>12.5 TeV.}
    \label{fig:corr}
\end{wrapfigure}
We estimated charge numbers of nuclei with two methods and Fig.\ref{fig:corr} shows the relation between the results from them. Naturally there is a positive correlation, though the correlation coefficient is small as $r=0.32$ because of broad charge resolution $\Delta Z\sim 6$ from the shower MVA.

\begin{figure}[t]
  \begin{center}
    \begin{minipage}[t]{7.0cm}
      \includegraphics[width=6.9cm,keepaspectratio]{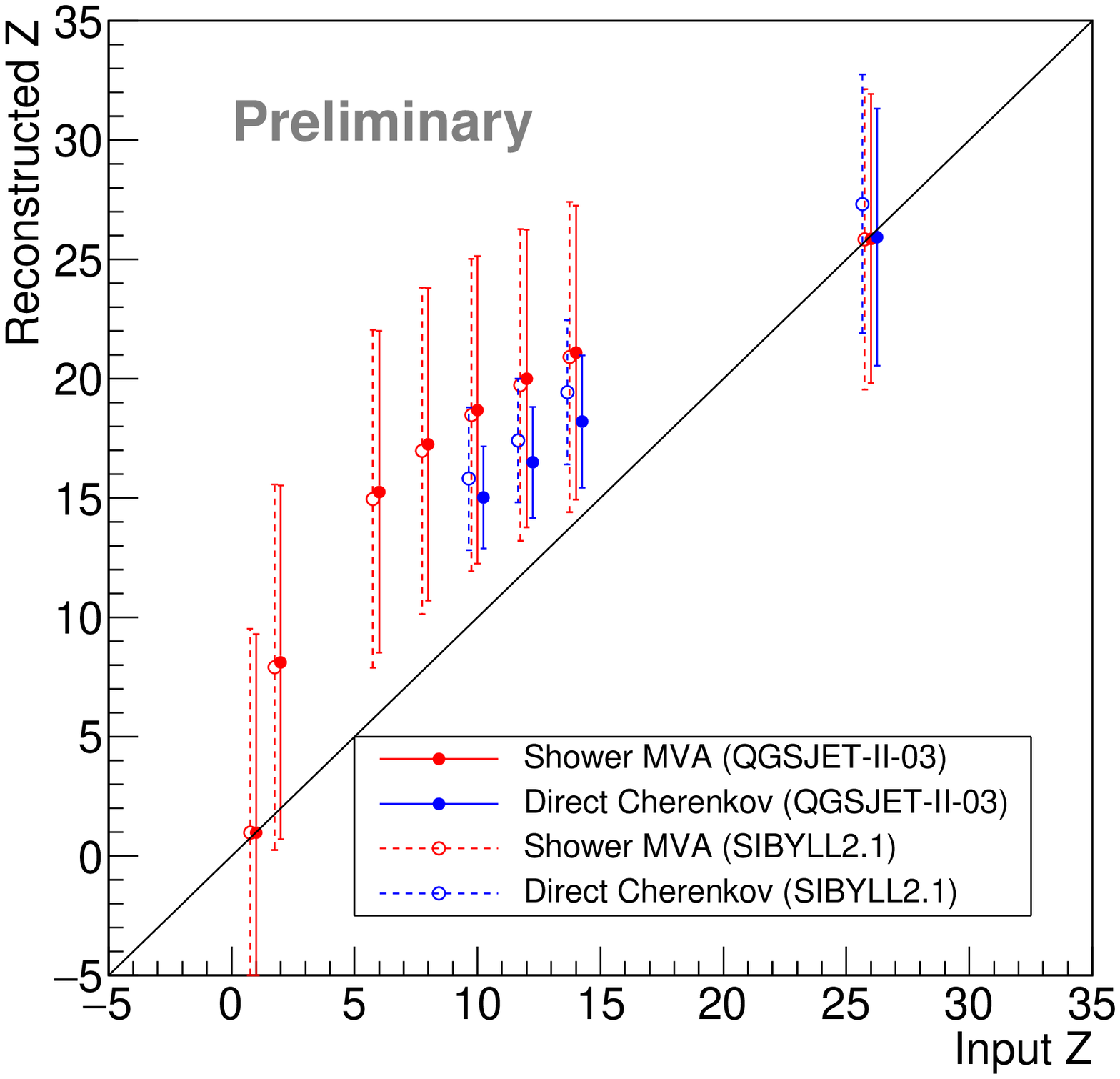}
      \caption{Relation between input charge and reconstructed charge from two reconstruction methods and two interaction models  ($E>12.5$ TeV, all events). Errorbars correspond to 1$\sigma$ obtained from Gaussian fit. Small offsets were added in x-axis direction for the visibility of the datapoints.}
      \label{fig:cwid_summary}
    \end{minipage}\hspace*{6mm}
     \begin{minipage}[t]{7.0cm}
       \includegraphics[width=6.9cm,keepaspectratio]{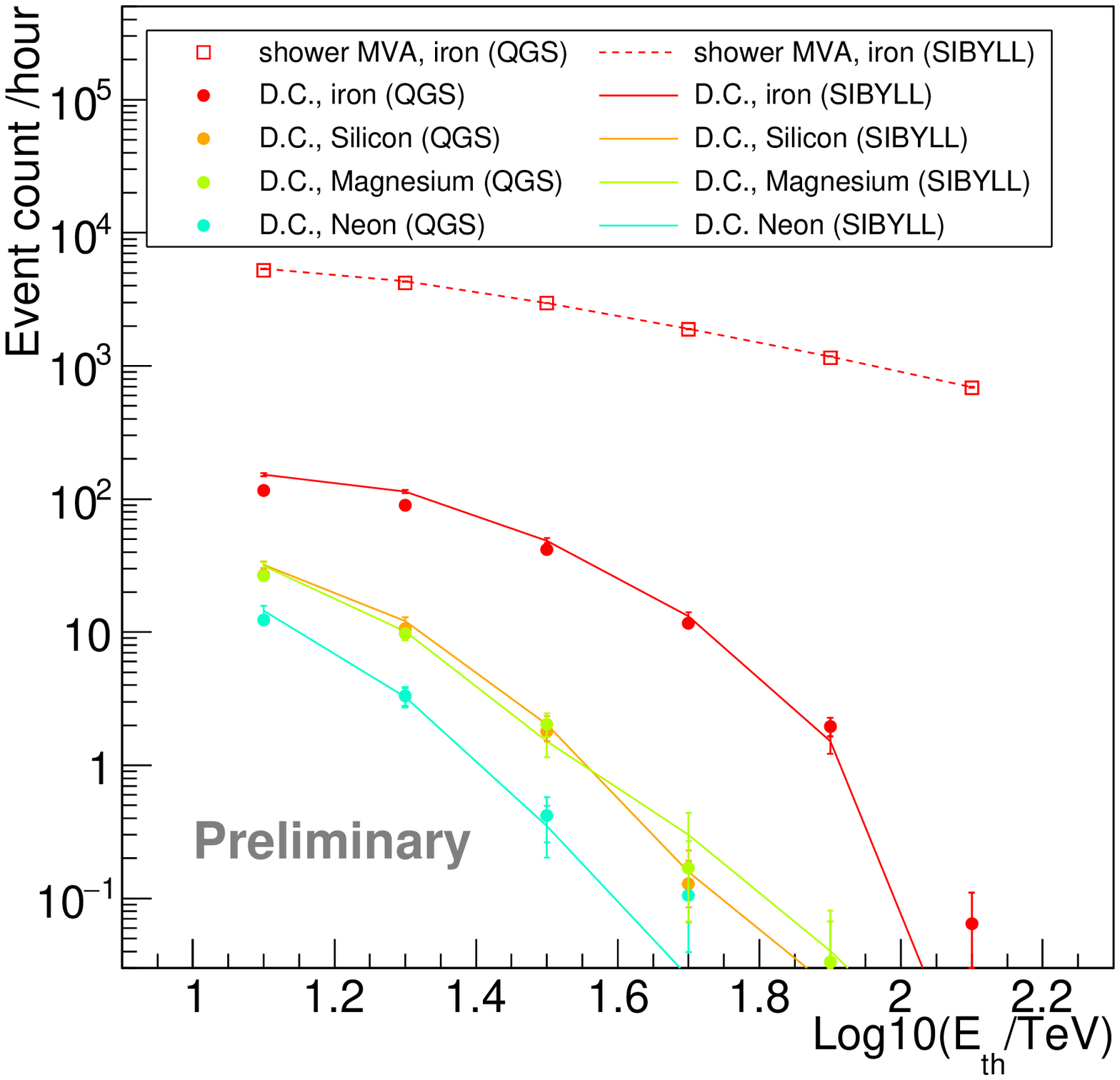}
       \caption{Expected event rates versus energy lower bound, with an assumption of cosmic-ray flux value from \cite{Horandel}. As for shower MVA, only iron (distant from broad proton peak) is plotted.}
       \label{fig:rate}
     \end{minipage}
    \end{center}
\end{figure}

Distribution widths of reconstructed charge obtained by Gaussian fit and expected count rates for 8 types of nuclei are summarized in Fig.\ref{fig:cwid_summary}, \ref{fig:rate} and Tab.\ref{tab:cwid_summary}. Note that distribution width strongly depends on energy both for DC and shower MVA. We applied identical analysis to QGSJET-II-03 and SIBYLL2.1 datasets and the results for both are shown. Difference of results between them (which is larger than statistical error) is regarded as a part of systematic errors.

We expect $\sim 180$ event counts per hour for a sum of Fe, Si, Mg and Ne from DC analysis in $E>12.5$ TeV region. As for DC, expected event rate decreases rapidly toward high energy since detection efficiency of DC event declines due to energetic bright shower in the background.
\begin{table}[t]
  \begin{center}
  \small
\begin{tabular}{c|cc|cc}
  \hline
  \multicolumn{5}{c}{Reconstructed charge width} \\
  \hline
   & \multicolumn{2}{|c|}{QGSJET-II-03}&\multicolumn{2}{|c}{SIBYLL2.1} \\
  Nucleus & Shower MVA & DC & Shower MVA & DC \\
  \hline
  Iron &  $6.05\pm0.02$ & $5.38\pm0.13$ & $6.30\pm 0.02$ & $5.41\pm0.12$\\
  Silicon & $6.16\pm0.03$ & $2.77\pm0.15$ & $6.50\pm 0.03$ & $3.02\pm0.13$
 \\
  Magnesium & $6.24\pm0.03$ & $2.33\pm0.14$ & $6.54\pm 0.03$ & $2.59\pm0.12$ 
 \\
  Neon & $6.44\pm0.03$  & $2.14\pm0.23$  & $6.55\pm 0.04$ & $2.99\pm0.25$
 \\
  Oxygen & $6.55\pm0.04$ & -& $6.84\pm0.04$ & - \\
  Carbon & $6.74\pm0.04$ & - & $7.08\pm0.04$ & -\\
  Helium & $7.41\pm0.03$ & -& $7.66\pm 0.03$ & -\\
  Proton & $8.30\pm0.03$ & - & $8.53\pm 0.03$ & -\\
  \hline
\end{tabular}
\caption{List of widths of reconstructed charge obtained by Gaussian fit for 8 types of nuclei (preliminary value). All events for $E>12.5$ TeV were used and only statistical errors are considered. As for direct Cherenkov analysis, values for lighter nuclei than oxygen are not shown because of poor event statistics.}
\label{tab:cwid_summary}
\end{center}
\end{table}

%% file: sec4.tex
We generated $\sim 10^{7}$ cosmic ray simulation events for 8 types of nuclei from H (Z=1) to Fe (Z=26) for CTA array which consists of 24 MSTs. Two types of charge reconstruction methods were applied and distribution widths of reconstructed charge (or charge resolution) and expected count rates assuming the literature flux value are summarized in Fig.\ref{fig:cwid_summary} and Fig.\ref{fig:rate}. As for iron (Z=26), charge resolution from shower MVA is approaching to that of direct Cherenkov within 17$\%$ and it would be helpful in high energy region where detection efficiency of DC events rapidly decreases. The expected count rate of iron obtained from shower MVA is $\sim 70$ times larger than that from DC analysis for $\log_{\rm 10}(E)>1.5$ region. As for medium weight nuclei (Ne-Si, Z=10-14), contribution of direct Cherenkov is still essential because of its high rejection power of proton and helium and better charge resolution ($\Delta Z\sim 3$).

 The analysis methods applied here were somewhat simplified and there is a significant room for the improvement. In order to treat two components (DC and shower) smartly, template likelihood fitting method \cite{TEMPLATE} in shower reconstruction and DC and shower combined MVA \cite{VERITAS} is a possible way to move forward. And other telescopes than MSTs will have significant power in cosmic-ray composition measurements, by the merit of isochronous reflector and fine pixel (direct Cherenkov analysis with LSTs) or large effective area in $E>100$ TeV region (shower MVA with SSTs).